\documentclass[12pt]{article}
\usepackage{sw20jart}

\input tcilatex
\QQQ{Language}{
British English
}

\begin{document}

\title{Entropic Principles}
\author{John D. Barrow \\
%EndAName
Astronomy Centre\\
University of Sussex\\
Brighton BN1 9QJ\\
UK}
\maketitle
\date{}

\begin{abstract}
We discuss the evolution of radiation and Bekenstein-Hawking entropies in
expanding isotropic universes. We establish a general relation which shows
why it is inevitable that there is currently a huge difference in the
numerical values of these two entropies. Some anthropic constraints on their
values are given and other aspects of the cosmological 'entropy gap' problem
are discussed. The coincidence of the classical and quantum entropies for
black holes with Hawking lifetime equal to the age of the universe, and
hence of radius equal to the proton size, is shown to be identical to the
condition that we obseve the universe at the main sequence lifetime.
\end{abstract}

\section{Introduction}

In this paper we discuss the inevitability of a number of simple relations
between the matter and radiation entropies of Friedmann universes and the
value of the Bekenstein-Hawking entropy of the visible universe. The huge
difference in values of these two possible entropies of the observable
universe has led some to the conclusion that the universe is in an
extraordinarily improbable state compared to one in which its radiation and
matter contents were reorganised. We will show that the entropy gap is a
consequence of the gravitational field equations alone and is just another
way of stating the cosmological 'flatness problem'. We also show how the
weak anthropic principle places very strong lower bounds on the entropy gap.
Since the dimensionless numbers involved in the resolution of these
questions are very large we shall provide an order of magnitude analysis
which ignores particle spin weight factors and numerical contributions $%
O(2\pi ).$ We begin by analysing the simplest situation of the flat,
radiation-dominated universe to establish some conclusions and then extend
the analysis to the open and matter plus radiation cases before making some
points about black hole entropy coincidences.

\section{Entropies in flat radiation universes}

Consider the Bekenstein-Hawking entropy, $S$, of the cosmological particle
horizon at comoving proper time, $t.$ It is given by the area of the horizon
size $\sim t$ in Planck units ($c=\hbar =k=1$); that is,

\begin{equation}
S\ \sim \frac A{A_p}\sim \left( \frac t{t_p}\right) ^2.  \label{a}
\end{equation}
The current value of this quantity, at $t=t_0\sim 10^{17}s$ $\sim 10^{60}t_p$
gives the cube of a Dirac large number:

\begin{equation}
S_0\sim \left( \frac{t_0}{t_p}\right) ^2\sim 10^{120}.  \label{b}
\end{equation}
We can regard $S$ as the maximum possible gravitational entropy for the
visible universe inside a Hubble radius of size $\sim t$. It is related to
the Bekenstein entropy bound \cite{bek}, that for any physical system of
size $R$ and energy $E$ the entropy is bounded above by

\begin{equation}
S\leq S_{\max }=ER.  \label{bek}
\end{equation}

This maximum value, $S_{\max },$ is equal to the value (\ref{a}) for a
gravitating system with $R\sim E/m_p^2\sim t.$ Now consider the radiation
entropy inside the horizon, $S_\gamma $. This is roughly equal to the number
of photons inside the horizon of size $\sim t$; thus, if the number density
of photons at temperature $T$ is $n_\gamma \sim T^3$, we have

\begin{equation}
S_\gamma \sim n_\gamma t^3\sim T^3t^3\sim \left( \frac T{T_p}\right)
^3(\frac t{t_p})^3.  \label{c}
\end{equation}
If the universe is flat, isotropic, and always radiation dominated then the
energy density of radiation, $\rho _\gamma $, is related to the comoving
proper time by a relation of the approximate form

\begin{equation}
\rho _\gamma \sim T^4\sim \frac 1{Gt^2}  \label{d}
\end{equation}
and so we always have

\begin{equation}
S_\gamma \sim \left( \frac t{t_p}\right) ^{3/2}.  \label{e}
\end{equation}
Hence today, at $t_0$, we have roughly

\begin{equation}
S_\gamma \sim \left( \frac{t_0}{t_p}\right) ^{3/2}\sim 10^{90}  \label{e1}
\end{equation}
and the explicit dependence of the values of $S$ and $S_\gamma $ on the time
when they are evaluated is clear.

\section{The Entropy Gap}

If we combine eqns. (\ref{a}) and (\ref{e}) we see that the following
relation between $S$ and $S_\gamma $ must hold at all times:

\begin{equation}
S\sim S_\gamma ^{4/3}.  \label{f}
\end{equation}
This relation is instructive. It shows that if we have a universe containing
radiation we are not at liberty to imagine that the radiation entropy could
be as large as $S$ except if the universe is of Planck size, in which case
we must have $S\sim S_\gamma \sim O(1).$ Even if explosive non-equilibrium
behaviour were suddenly to erupt in the Universe, the eventual equilibrium
state would have to satisfy (\ref{f}) if the expansion were still isotropic.
This observation is relevant to the argument of Penrose \cite{pen} that the
present state of the universe is highly improbable because the maximum
entropy that it could have, $S(t_0)$, is so much larger than the observed
radiation entropy, $S_\gamma (t_0)$. This argument implies that the matter
could be rearranged to make the radiation or matter entropy as large as $%
S\sim 10^{120}$ today without changing $S$. However this is clearly not the
case. If anything were done to raise the radiation entropy to a value of
order $S(t_0),$ then Einstein's equations would couple the new radiation
density to the age and size of the horizon via eq. (\ref{d}) and we would
have to have $S\sim 10^{160}$. In this respect, the gap between $S$ and $%
S_\gamma $ tells us nothing about the probability or improbability of the
present state of the observable universe or of its initial data. The $%
S-S_\gamma $ gap is a direct consequence of the law of gravitation.
Moreover, we see that the vast difference in the values of $S$ and $S_\gamma 
$ actually follows from an initial state at the Planck scale where $S_\gamma
\sim S$ which would have to be judged highly probable if the same reasoning
were applied to it by 'observers' existing at that time.

A similar relation to that given by (\ref{f}) also applies to oscillating
closed universes, with the entropy values evaluated at the expansion maxima
of successive cycles. It shows very simply why an increase in thermal
radiation entropy ($S_\gamma $) from cycle to cycle would require an
increase in the size of each cycle's expansion maximum (determined by $S$)
and hence a closer approach of the closed universe to 'flatness'. This was
first pointed out by Tolman \cite{tol} and recently generalised by Barrow
and Dabrowski \cite{bd}. It is interesting to note that the presence of a
positive cosmological constant always requires these oscillations to cease
and be replaced by expansion towards a de Sitter state.

We note that the 'entropy gap' between $S$ and $S_\gamma $ is also the
reason why the traditional 'heat death' of the universe does not occur in
ever-expanding isotropic universes \cite{bt}. There is no approach to
gravitational equilibrium. Although the radiation entropy increases (and may
be augmented by other non-equilibrium processes) in accord the expectations
of the second law, the maximum entropy defined by the gravitational horizon
entropy $S$ increases faster and so the entropy gap $S-S_{\gamma \text{ }}$%
grows with time: the universe gets\textit{\ farther }from the equilibrium of
Helmholtz's 'heat death' \cite{heat}, \cite{bt}, since $S/S_\gamma \sim
(t/t_p)^{1/2}$ as $t\rightarrow \infty $. In fact, in more general
anisotropic and inhomogeneous universes the 'heat death' is a more
complicated question that involves the analysis of the asymptotic evolution
of the anisotropic gravitational-wave distortions to the expansion dynamics.
It appears that there need be no gravitational 'heat death' either, unless
there is a positive cosmological constant \cite{bt}.

We should remark that the total entropy of the Universe may not be a finite
quantity. Without delving into the possible classical, dynamical, quantum
gravitational, or stringy contributions to the total entropy we can see that
if the Universe is infinite in volume and contains a uniform distribution of
black holes then the total Bekenstein-Hawking entropy of the black holes
will be infinite. Any evaluations of the total entropy of the Universe are
therefore problematic in the infinite volume case, as are all other global
aspects of such universes \cite{imposs}.

There is a clear anthropic aspect. If observers who evolve by natural
selection can only be on the cosmic scene after a time of order the
main-sequence stellar lifetime for hydrogen burning, $t_{ms}\sim
m_p^2m_N^{-3},$ then we must be observing the universe at a value of $t_0$
that is bounded by \cite{dicke}, \cite{carr},

\begin{equation}
t_0>t_{ms}\sim \left( \frac{m_p}{m_N}\right) ^2m_N^{-1}\sim 10^{38}\times
10^{-23}s\sim 10^9yrs,  \label{h}
\end{equation}
where $m_N=1.67\times 10^{-24}gm$ is the proton mass. If we fail to find
ways of existing after the stars have died then we may be more strongly
constrained to observe only when $t_0\sim t_{ms}.$ But assuming only (\ref{h}%
), we would have to find values of $S$ and $S_\gamma $ which satisfy

\begin{eqnarray}
S &\geq &\left( \frac{t_{ms}}{t_p}\right) ^2\sim \left( \frac{m_p}{m_N}%
\right) ^6\sim 10^{114},  \label{anth1} \\
S_\gamma &\geq &\left( \frac{t_{ms}}{t_p}\right) ^{3/2}\sim \left( \frac{m_p%
}{m_N}\right) ^{9/2}\ \sim 10^{86}.  \label{anth2}
\end{eqnarray}

Our existence would not be possible if the values of $S$ and $S_\gamma $
were significantly smaller. A slightly weaker (but more fundamental)
anthropic constraint can be derived by arguing that observers can only exist
when the temperature falls below the ionisation energy of atoms; that is,
when the radiation temperature satisfies

\begin{equation}
T<T_{ion}\sim \alpha ^2m_e,  \label{an3}
\end{equation}
where $\alpha =1/137.04$ is the fine structure constant and $m_e$ $%
=9.1\times 10^{-28}gm$ is the electron mass. Using (\ref{d}), this means
that we must be observing the universe when its age satisfies

\[
t^{\ }>t_{ion}\sim \frac{m_p}{\ T_{ion}^2}\sim \alpha ^{-4}\left( \frac{m_p}{%
m_e}\right) ^2t_p\sim \ 10^{12}\ s\ , 
\]
which would require atom-based observers to witness

\begin{eqnarray}
S &\geq &\left( \frac{t_{ion}}{t_p}\right) ^2\sim \alpha ^{-8}\left( \frac{%
m_p}{m_e}\right) ^4\sim 10^{110},  \label{anth4} \\
S_\gamma &\geq &\left( \frac{t_{ion}}{t_p}\right) ^{3/2}\sim \alpha
^{-6}\left( \frac{m_p}{m_e}\right) ^3\ \sim 10^{83}.  \label{anth5}
\end{eqnarray}
If we were to require the observers to have molecular structure then the
ionisation temperature would be replaced by the dissociation temperature and
the lower bounds on the entropies would only be slightly increased by a
geometrical factor.

\section{Open universes}

So far, for simplicity, we have assumed that the universe is spatially flat
and contains only radiation. This is a good approximation to the actual
state of affairs because so many decades of evolution occurred in the
radiation era. But we can easily generalise our results to the cases where
these assumptions do not hold. Consider first the situation of an open
radiation universe. To an excellent approximation the dynamics are described
by an expansion scale factor $a(t)\propto t^{1/2}\propto T^{-1}$ for the
flat universe up until some characteristic time $t_c$, after which the
expansion is curvature dominated with $a(t)\propto t\propto T^{-1}$. Hence,
the maximum entropy, $S$, inside a scale of size $\sim t$ is still given by (%
\ref{a}), but the radiation entropy is given by (\ref{c}) or (\ref{e}) only
at times $t\leq t_c$, before the universe becomes curvature dominated. At
later times we have

\begin{equation}
S_\gamma \sim \left( \frac{t_c}{t_p}\right) ^{3/2};t\geq t_c,  \label{op}
\end{equation}
\textit{independent of time }$t.$ Thus $S_{\gamma \text{ }}$remains of order
the value it had at the time when the expansion first became curvature
dominated. The earlier results for the flat universe are obviously recovered
by putting $t_c=t_0$. Thus we have

\begin{equation}
\frac S{S_\gamma }\sim \left( \frac t{t_p}\right) ^2\left( \frac{t_p}{t_c}%
\right) ^{3/2};t\geq t_c.  \label{op2}
\end{equation}
Again, we see there is no heat death: the entropy gap grows even faster than
in the flat universe as $t\rightarrow \infty $. As in the flat case, we see
that the fact that we observe $S>>S_\gamma $ is a direct consequence of the
cosmological evolution equations and nothing to do with the likelihood of
particular initial conditions or the present state of the visible universe.
In order to have a universe with $S\sim S_\gamma $ we would need it to have
begun expanding right from the quantum era in a curvature-dominated state,
that is $t_c\sim t_p,$ and to observe it at a time $t_0\sim t_p$. This is a
situation that cannot be observed by beings like ourselves. The same general
conclusions hold regarding anthropic constraints on the observed values of $%
S $ and $S_\gamma $ and we can supplement them with the requirement that we
need $t_c>t_{ms}$ if stars are to form by the process of gravitational
instability from small primordial density perturbations \cite{cart, jdb}.

The relations (\ref{a}), (\ref{e}), and (\ref{f}) also reveal the source of
the large present-day values today of $S$ and $S_\gamma $. The large numbers
reflect the large value of the present age of the universe, $t_0$, in Planck
units. Thus the largeness of $S$ and $S\gamma $ is a way of restating the
flatness problem of cosmology for which inflation provides a possible
solution \cite{guth} by supplying a mechanism for enlarging the size of a
universe that begins expanding with $t_c$ very close to $t_p$.

\section{Matter and Radiation Entropies}

Now consider the addition of matter to the flat radiation universe. There
will appear a characteristic time, $t_{eq},$ determined by the relative
number densities of protons and photons such that at times earlier than $%
t_{eq}$, the dynamics are radiation dominated and eqns. (\ref{a})-(\ref{e})
apply; but in the matter-dominated era after $t_{eq}$ the expansion scale
factor evolves as $a(t)\propto t^{2/3}.$ Thus $S$ will be given by (\ref{a})
and $S_\gamma $ by

\begin{equation}
S_\gamma \sim \left( \frac{t_{eq}}{t_p}\right) ^{1/2}\left( \frac
t{t_p}\right) ;t\geq t_{eq}.  \label{eq1}
\end{equation}
In general, therefore, we have a simple relation between $S_\gamma $ and $S:$

\begin{equation}
S_\gamma \sim \left( \frac{t_{eq}}{t_p}\right) ^{1/2}\ S^{1/2};t\geq t_{eq}
\label{eq2}
\end{equation}
The previous results for the flat radiation universe are recovered by
putting $t_{eq}\sim t_0.$ In our universe we have a matter-radiation balance
which implies that $t_{eq}\sim 10^{10}s\sim 10^{53}t_p.$

It is easy to see what will happen in the most general case of an open
universe containing matter and radiation. The gravitational entropy will
always obey (\ref{a}) but the radiation entropy will have a value today $($%
where $t_0>t_c)$ given by (\ref{eq2}) evaluated at $t=t_c,$ that is by

\begin{equation}
S_\gamma \sim \left( \frac{t_{eq}}{t_p}\right) ^{1/2}\left( \frac{t_c}{t_p}%
\right) ;t\geq t_c\geq t_{eq}  \label{op3}
\end{equation}

We recover the radiation-universe results if we put $t_0=t_{eq}$ in (\ref
{op3}). Again, we see that there is a gravitationally determined link
between the values of $S$ and $S_\gamma .$ In order for them to be similar
in magnitude we would require the universe to have $t_0^2\sim
t_p^{1/2}t_ct_{eq}^{1/2}.$

We can also consider the fate of the classical matter entropy, $S_m$, which
is determined by the total number of particles. If we count particles of
mass $m$ inside the scale $t,$ then

\[
S_m\sim n_mt^3\ \propto a^{-3}t^2. 
\]
Hence, in a flat matter-dominated evolution we have $S_m\sim $ constant, but
in an open curvature-dominated universe we approach a zero-entropy asymptote
with $S_m\propto t^{-1}.$ We may also write

\[
S_m\sim n_mt^3\sim \frac{\rho t^3}m\sim \left( \frac t{t_p}\right) \left( 
\frac{m_p}m\right) 
\]
so the observed entropy per proton is given in the flat case (or at $t_0<t_c$
in an open universe) by

\[
\frac{S_\gamma }{S_m}\sim \left( \frac{t_0}{t_p}\right) ^{1/2}\left( \frac
m{m_p}\right) \sim 10^9. 
\]

If we determine the classical entropy by counting particles other than
protons then we can alter the absolute numerical value. Conventionally, we
count protons, but if the universe is dominated by non-baryonic dark matter
particles --- for example by the lightest supersymmetric particle or by
axions --- then they may be more appropriate entropy counters and the
particle-entropy numerology will be changed.

\section{A Black Hole Coincidence}

We may also determine classical and quantum (Bekenstein-Hawking) entropies
for a Schwarzschild black hole of mass $M$ and radius $R_{BH}=2Mm_p^{-2}$.
>From (\ref{a}), the quantum entropy of the black hole is

\[
S_{BH}\sim \left( \frac M{m_p}\right) ^2 
\]
and its Hawking lifetime \cite{swh} is given by

\[
t_{BH}\sim \frac{M^3}{m_p^4}. 
\]
The classical entropy of the black hole, $S_{cl},$ is determined by the
number of particles of mass $m$ it 'contains'. This is given by

\[
S_{cl}\sim \frac Mm. 
\]

We see that there is an interesting coincidence \cite{dws}: black holes
which have a Hawking lifetime equal to the present age of the universe, $%
t_{BH}\sim t_0$, have classical and quantum entropies that are similar in
value (if $m\sim m_N\sim 10^{-24}gm$) and a Schwarzschild radius equal to
the proton radius $\sim m^{-1}$ because

\[
\frac{S_{BH}}{S_{cl}}\sim \left( \frac m{m_p}\right) \left( \frac{t_p}{t_{BH}%
}\right) ^{1/3}\ \sim \frac{Mm}{m_p^2}. 
\]
However, we note that the coincidence that $S_{BH}\sim S_{cl}$ for $%
t_{BH}\sim t_0$, reduces to the condition that the present epoch is

\[
t_0\sim \frac{m_{p\text{ }}^2}{m^3}. 
\]
By reference to eq. (\ref{h}), we see that this is precisely the condition
that we are living at about the main-sequence age: $t_0\sim t_{ms}.$ If we
chose to count the classical entropy of the black hole using a reference
particle mass, $m,$ not equal to the proton mass then this coincidence would
no longer hold but the fact that for black holes with $S_{BH}\sim S_{cl}$ we
must have $R_{BH}\sim m^{-1}$ would still hold regardless of the identity of
the particle with mass $m$.

\section{Conclusions}

We have shown that various measures of the entropy of the observable
universe can be estimated by simple order-of-magnitude analysis. This
reveals the necessary connection between the radiation and matter entropies
and the Bekenstein-Hawking (BH) entropy of the mass density contained within
the Hubble radius at any time. This BH entropy is generally taken to define
the maximum entropy that the observable universe could possess. The
smallness of the radiation entropy with respect to the BH entropy is shown
to be a necessary consequence of the Friedmann equations governing the
expansion of the universe. Its relative smallness should not therefore be
interpreted as telling us that the material in the observable universe is
(or was) in an extraordinarily low entropy state. The entropy gap between
the radiation entropy and the BH entropy necessarily grows with time and the
magnitude of the gap is a way of measuring the size and age of the
observable universe. The gap could only be negligible in a universe that was
too young and dense for living observers to exist: it is another way of
stating the flatness problem.

\textbf{Acknowledgement}

The author is supported by a PPARC Senior Fellowship.

\end{document}